\documentclass[aps,floats,twocolumn,showpacs,showkeys]{revtex4}
\usepackage{graphicx}
\usepackage{epsfig}
\input{psfig.sty}

\newcommand{\bastar}{\begin{eqnarray*}}
\newcommand{\eastar}{\end{eqnarray*}}
\newskip\humongous \humongous=0pt plus 1000pt minus 1000pt

\newif\ifdtup

\relax
\newcommand{\be}{\begin{equation}}
\newcommand{\ee}{\end{equation}}
\newcommand{\bea}{\begin{eqnarray}}
\newcommand{\eea}{\end{eqnarray}}
\newcommand{\pro}{\partial}
\newcommand{\n}{\hat n}
\newcommand{\oneg}{\displaystyle\frac{1}{g}}
\newcommand{\D}{{\hat D}}
\newcommand{\X}{{\vec X}}
\newcommand{\vX}{{\vec X}}

\newcommand{\tC}{{\tilde C}}
\newcommand{\tD}{\tilde D}
\newcommand{\dfrac}{\displaystyle\frac}
\newcommand{\ba}{\begin{array}}
\newcommand{\ea}{\end{array}}
\newcommand{\nn}{\nonumber}

\begin{document}
\title{Stable monopole-antimonopole string background in $SU(2)$ QCD}
\bigskip
\author{Y. M. Cho $ ^{a,b}$}
\email{ymcho@yongmin.snu.ac.kr}
\author{D. G. Pak $ ^{c}$}
\affiliation{ $ ^a $
C. N. Yang Institute for Theoretical Physics,State University of New York,
 Stony Brook, New York 11794, USA \\
$ ^b$ School of Physics, College of Natural Sciences,
Seoul  National University, Seoul 151-742, Korea\\
$ ^c $ Center for Theoretical Physics,
Seoul National University, Seoul 151-742, Korea}
\email{dmipak@phya.snu.ac.kr}
\begin{abstract}
Motivated by the instability of the Savvidy-Nielsen-Olesen (SNO)
vacuum we make a systematic search for a stable magnetic background
in pure $SU(2)$ QCD. It is shown that a pair of axially symmetric 
monopole and antimonopole strings is stable, provided that the 
distance between the two strings is less than a critical value.
The existence of a stable monopole-antimonopole string
background strongly supports that a magnetic condensation
of monopole-antimonopole pairs can generate
a dynamical symmetry breaking, and thus
the magnetic confinement of color in QCD.
\end{abstract}
\maketitle
{\bf 1. Introduction}

It has long been argued that the
monopole condensation could explain the confinement of color
through the dual Meissner effect \cite{nambu}. Indeed, if one
assumes the monopole condensation, one could easily argue that the
ensuing dual Meissner effect guarantees the confinement
\cite{cho1}. There have been many attempts to prove this scenario
in QCD \cite{savv,cox}. Unfortunately, the earlier attempts had
encountered the well known problem of Savvidy-Nielsen-Olesen (SNO)
vacuum instability \cite{savv}. In fact the effective action of
QCD obtained with the SNO vacuum develops an imaginary part, which
implies that the vacuum is unstable \cite{niel2,rajiad}. This
instability of the magnetic condensation has been widely accepted
and never been convincingly revoked.

Recently it has been shown that, if one uses
a proper infra-red regularization which respects causality,
the imaginary part in the effective action
can be removed \cite{cho3}. In addition,
a numerical evidence for the stable magnetic
condensation has been found in lattice simulation
\cite{kondo1}.

The purpose of this paper is to search for a stable classical
magnetic background in $SU(2)$ model of QCD.
We analyze the stability of two classical magnetic
backgrounds, a pair of axially symmetric monopole-antimonopole
strings and a pair of magnetic vortex-antivortex strings,
and show that the first one becomes stable
provided the distance between two strings is less
than a critical value.
As far as we understand, the pair of axially symmetric
monopole-antimonopole strings constitutes a first explicit example
of a stable magnetic background in QCD.
More importantly, the result can serve as a strong
indication that a monopole-antimonopole condensation
can provide a stable vacuum in QCD.

{\bf 2. Instability of Wu-Yang monopole background}

Let us start with a gauge invariant Abelian projection in
$SU(2)$ model of QCD which includes explicitly a topological degree
of freedom expressed by the unit isotriplet $\hat n$.
We decompose the gauge potential into the restricted
potential $\hat A_\mu$ and the off-diagonal part (valence gluon) $\vec X_\mu$
\cite{cho1}
\bea
&& \vec{A}_\mu =A_\mu \n - \oneg \n\times\pro_\mu\n+\X_\mu\nonumber
= \hat A_\mu + \X_\mu, \nn\\
&&(\n^2 =1,~~~ \hat{n}\cdot\vec{X}_\mu=0),
\label{cdec}
\eea
Notice that the restricted potential is precisely the connection which
leaves $\n$ invariant under the parallel transport,
\bea
&&\D_\mu \n = \pro_\mu \n + g {\hat A}_\mu \times \n = 0.
\eea
The restricted potential $\hat{A}_\mu$ has a dual structure which
can be seen from the field strength decomposition
\begin{eqnarray}
&\hat{F}_{\mu\nu}=(F_{\mu\nu}+ H_{\mu\nu})\hat{n}\mbox{,}\nonumber\\
&F_{\mu\nu}=\partial_\mu A_{\nu}-\partial_{\nu}A_\mu, \nn\\
&H_{\mu\nu}=-\dfrac{1}{g} \hat{n}\cdot
(\partial_\mu\hat{n}\times\partial_\nu\hat{n})
=\partial_\mu \tilde C_\nu-\partial_\nu \tilde C_\mu,
\end{eqnarray}
where $\tilde C_\mu$ is the ``magnetic'' potential
\cite{cho1}.

With the decomposition (\ref{cdec}) one has
\bea
\vec{F}_{\mu\nu}&=&\hat F_{\mu \nu} + \D _\mu \vX_\nu -
\D_\nu \vX_\mu + g\vX_\mu \times \vX_\nu,
\eea
so that the Lagrangian can be written as follows
\bea
&{\cal L} =-\dfrac{1}{4} {\hat F}_{\mu\nu}^2
-\dfrac{1}{4}(\D_\mu\vX_\nu-\D_\nu\vX_\mu)^2 \nn \\
&-\dfrac{g}{2} {\hat F}_{\mu\nu} \cdot (\vX_\mu \times \vX_\nu)
-\dfrac{g^2}{4} (\vX_\mu \times \vX_\nu)^2.
\eea
With the gauge fixing condition $\hat D_\mu \vX^\mu=0$
the one-loop correction to the effective action reduces to the form
\bea
&&\exp {(i\Delta S)}={\rm Det}^{-1} K_{\mu\nu}
~{\rm Det}^2 M_{FP}, \nn \\
&& K_{\mu\nu}=g_{\mu\nu} \tD \tD + 2 i (F_{\mu\nu}+H_{\mu\nu}), \nn \\
&& M_{FP} =\tD \tD ,  \,\,\,\,\,\,\,
     \tD_\mu =\partial_\mu + ig (A_\mu+\tilde{C}_\mu) ,  \label{effaction0}
\eea
where the operators $K_{\mu\nu}$ and $M_{FP}$
originate from the functional integration over the off-diagonal
gluon and Faddeev-Popov ghost respectively (the contribution from integration
over the quantum part of the Abelian field $A_\mu$ is trivial).
For arbitrary static magnetic background $F_{\mu\nu}+H_{\mu\nu}$
one can simplify the one-loop correction
\cite{cho3}
\bea
&&\Delta S = i\ln {\rm Det}(-\tD^2 + 2a)
         + i\ln {\rm Det} (-\tD^2 -2a), \label{dets}
\eea
where $a = g \sqrt{\dfrac{1}{2} H_{\mu\nu}^2}$,
hereafter, for the brevity of notation, we employ a single
notation $H_{\mu\nu}$ for the additive combination
$F_{\mu\nu}+H_{\mu\nu}$.

Before we discuss the stability of monopole-antimon-\newline
opole pair, we first review the instability of the Wu-Yang
monopole because two problems are closely related \cite{brandt}.
The Wu-Yang monopole solution
of charge $q/g$ is described by \cite{wu1,cho80}
\bea
&\vec A_\mu=-\dfrac{1}{g} \n \times \pro_\mu \n, \nn\\
&\n = \Bigg(\matrix{\sin \theta \cos{q\phi} \cr
\sin \theta \sin{q\phi} \cr
\cos \theta}\Bigg),
\label{mono}
\eea
where $(r,\theta,\phi)$ is the spherical coordinates
and $q$ is an integer monopole charge.
In the Abelian formalism it is more convenient
to describe the monopole in terms of the magnetic potential
$\tilde C_\mu$,
This implies the
components of the magnetic potential
$\tilde C_\mu$ and the magnetic field strength $H$
to be as follows
\bea
&& \tC_\mu = \dfrac{q}{gr} (\cos \theta-1) \pro_\mu \phi, \nn \\
&& H_{ij} = \dfrac{q}{g} \epsilon_{aij} \dfrac{x^a}{r^3} \label{paramcho}.
\eea
We use the parametrization for the magnetic potential \cite{cho1}
slightly different  from the parametrization in \cite{wu2} where
the magnetic potential  is defined on two coordinate
patches.

To study the stability of the monopole background
we should consider an operator obtained
by taking the second variation of the classical Lagrangian
with respect to
small fluctuations of the field $\vX_\mu$.
The operator is identical to the operator
$K_{\mu\nu}$, (\ref{effaction0}), and the
problem of finding unstable modes is reduced to
calculation of the scalar functional determinants in (\ref{dets})
\bea
&{\rm Det} \, K = {\rm Det}(-\tilde D^2 \pm 2a),
\eea
here, $a=\dfrac{q}{r^2}$ is given by
the magnetic field strength of the Wu-Yang monopole.
The absence or presence of negative modes of the operator $K$
implies stability or instability of the classical
background against small fluctuations of the gauge potential.
To calculate the eigenvalues of the operator $K$ one can write down
the eigenvalue equation as the following Schroedinger type equation
for a complex scalar field $\Psi$ which plays a role of wave function
\bea
&&K \Psi(r,\theta, \phi) = E \Psi(r, \theta, \phi),  \nn\\
K &=& -\Delta - 2\dfrac{iq}{r^2\sin^2 \theta} \cos \theta \pro_\phi
+ \dfrac{q^2}{r^2} \cot^2 \theta \pm 2\dfrac{q}{r^2}, \nn \\
\Delta &=&\dfrac{1}{r^2}\pro_r( r^2 \pro_r) + \dfrac{1}{r^2 \sin \theta}
\pro_\theta (\sin \theta \pro_\theta) + \dfrac{1}{r^2 \sin^2 \theta}
\pro^2_\phi \nn \\
&\equiv&\dfrac{1}{r^2}\pro_r( r^2 \pro_r) - \dfrac{\hat L^2}{r^2},
\label{schr}
\eea
where $\hat L$ is the angular momentum operator.
Notice that here the $\pm$ signatures represent
two spin orientations of the valence gluon.

Substituting the separated solution 
\bea
\Psi(r, \theta, \phi)= R(r) Y(\theta,\phi),
\eea
into (\ref{schr}) one obtains the equation for the angular eigenfunction
$Y(\theta,\phi)$
\bea
&\Big (\hat L^2 - \dfrac{2iq\cos \theta}{\sin^2 \theta} \pro_\phi
+ q^2 \cot^2 \theta  \Big) Y(\theta,\phi)  \nn\\
&= \lambda Y(\theta, \phi).
\label{mhe}
\eea
Moreover, with
\bea
&& Y(\theta,\phi) =  \dfrac{}{}\sum_{m=-\infty}^{+\infty}
\Theta_m (\theta) \Phi_m (\phi), \nn \\
&& \Phi_m (\phi) = \dfrac{1}{\sqrt {2\pi}} \exp(im\phi).
\eea
one can reduce (\ref{mhe}) to
\bea
\Big(-\dfrac{1}{\sin \theta} \pro_\theta (\sin \theta \pro_\theta)
+ \dfrac{(m +q \cos \theta)^2 }{\sin^2 \theta} \Big)\Theta
= \lambda \Theta.
\eea
This is exactly the eigenvalue equation for the monopole
harmonics which has been well-studied in the literature \cite{wu1,wu2}.
From the equation one obtains the following expression
for the monopole harmonics and the corresponding
eigenvalue spectrum
\bea
&& Y_{qjm}(\theta,\phi) =
\Theta_{qjm}(\theta) \Phi_m (\phi), \nn \\
&&\Theta_{qjm}(\theta) = (1-\cos \theta)^{\gamma_+}
(1+ \cos \theta)^{\gamma_-} P_k (\cos \theta), \nn\\
&& \gamma_\pm=\dfrac{|m \pm q|}{2}, \nn\\
&&\lambda = j(j +1) -q^2, \nn \\
&&j = k+\gamma_+ + \gamma_- , ~~~~~~~~~k=0,1,2,...,
\label{qnumber}
\eea
where $P_k(x)$ is the Legendre polynomial of order $k$.
The quantum number $j$ is analogous to the orbital
angular momentum quantum number $l$ of the standard spherical
harmonics $Y_{lm}$, except that here $j$ starts from
a non-zero integer value for a non-vanishing monopole
charge $q$.

Now, consider the equation for the radial
eigenfunction
\bea
&\Big(\dfrac{1}{r^2} \dfrac{d}{dr} ( r^2 \dfrac{d}{dr})
-\dfrac {1}{r^2} \big[j(j +1) -q^2\big]
\mp \dfrac{2q}{r^2} +E \Big)R(r) \nn \\
& = 0.
\eea
With $R(r)=\dfrac{1}{r}\chi(r)$ one obtains
\bea
\Big(\dfrac{d^2}{dr^2} - \dfrac{1}{r^2}\big[j(j+1) - q^2 \pm 2 q \big]
+ E \Big) \chi(r) =0.
\label{eqnchi}
\eea
The equation has a general solution in terms of Bessel functions
of the first kind $J_\nu (z)$
\bea
\chi (r) &=& \sqrt r \Big[C_1 J_{-\nu} (\sqrt E r) + C_2
J_\nu (\sqrt E r) \Big], \nn \\
 \nu &=& \dfrac{1}{2} \sqrt {1 + 4 [j(j+1) - q^2 \pm 2 q]},
\eea
where $C_i~(i=1,2)$ are the integration constants.
For positive values of $\nu$ and $E$ the finite solutions
oscillating at the infinity and vanishing at the origin
are given by $C_1 =0$.
The negative eigenvalues of $E$ can come
only from (\ref{eqnchi})
with the lower negative sign (which corresponds
to the operator $-\tilde D^2 - 2q/r^2$) and the lowest value of
$j=1$ with $q=1$. In this case we have to solve the equation
\bea
\Big(\dfrac{d^2}{dr^2} + \dfrac{1}{r^2} + E \Big) \chi =0,
\label{chi}
\eea
which is nothing but the one-dimensional
Schroedinger equation with the attractive
potential $-1/r^2$ \cite{landau}. The solution to this equation 
has a continuous eigenvalue spectrum 
for both positive and negative energies, and leads to
the radial eigenfunction $R(r)$ 
which behaves like
\bea
R(r)\simeq \dfrac{\sin \log (\sqrt {|E|} r)+ {\rm const}}{\sqrt r}
\eea
near the origin.
The solution has
an infinite number of zeros approaching the point $r=0$.
This can be interpreted as a valence gluon
moving around the monopole and falling down
to the center \cite{landau,brandt}.

One can observe that for the lowest energy states, ($j=1$),
the undesired attractive
potential proportional to $-1/r^2$ in (\ref{eqnchi}) vanishes
when $q=0$.
This can serve as a hint that one might expect the absence of
negative modes for a magnetic background
with zero monopole charge.
The simplest magnetic configuration with a total vanishing monopole charge
can be realized as a Wu-Yang monopole-antimonopole pair.
Unfortunately, since near the location of the (anti-)monopole
we still have the attractive potential part $-1/r^2$ one can verify
that the Wu-Yang monopole-antimonopole pair has to be unstable.

{\bf 3. Axially symmetric monopole string}

The axially symmetric monopole string
can be regarded as an infinite string carrying homogeneous
monopole charge density along the string.
The magnetic field strength of the axially symmetric
monopole string can be written in a simple form in the cylindrical
coordinates $(\rho,\phi,z)$
\bea
&&\tC_\mu=0, ~~~~~ A_\mu = -\alpha (z + \tau) \pro_\mu \phi, \nn \\
&&\vec H = \dfrac {\alpha}{\rho} \hat \rho, \label{mstring}
\eea
where $\alpha$ is the monopole charge density and
$\tau$ is an arbitrary constant which represents the translational
invariance of the magnetic field along the $z$-axis.
Just like the monopole solution (\ref{mono}) the above
monopole string forms a singular classical solution of $SU(2)$ QCD.
In the following we will assume $\alpha=1$ and $\tau=0$
for simplicity, since this will not affect the stability
analysis.

Let us consider the eigenvalue problem for the operator $K$ \bea
K\Psi(\rho,\phi,z) = E \Psi(\rho,\phi,z). \label{mseveq} \eea With
\bea \Psi=\dfrac{}{}\sum_{m=-\infty}^{+\infty}F_m(\rho,z)
\Phi_m(\phi), \eea and repeating the steps of the previous section
we obtain the following eigenvalue equation \bea F_{\rho \rho} +
\dfrac{1}{\rho} F_\rho + F_{zz} - \Big[\dfrac{(m-z)^2}{\rho^2} \pm
\dfrac{2}{\rho} - E \Big] F =0. \label{1aximon} \eea By shifting
$z$ to $z + m$ one can put $m=0$. The quantum mechanical potential
of this equation behaves like $\pm 2/\rho$ near $\rho=0$. So we
still have an undesired attractive potential $-2/\rho$. This
implies two things. First, the attractive interaction of the
axially symmetric monopole string background is less severe than
the attractive interaction of the spherically symmetric monopole
background. So we can expect the absence of continuous negative
energy spectrum for the axially symmetric monopole string
background. Secondly, the attractive potential $-2/\rho$ tells
that the monopole string background must still be unstable,
because it indicates the existence of discrete bound states with
negative energy.

To confirm this we make a qualitative estimate
of the negative energy eigenvalues of (\ref{1aximon}).
We look for a solution which has the form
\bea
&F(\rho,z) = \dfrac{}{} \sum_{n=0}^\infty f_n(\rho)
Z_n(x), \nn\\
&Z_n (x) = \exp (-\dfrac{x^2}{2}) H_n(x), \nn\\
&x = \dfrac{z}{\sqrt \rho},
\label{eqnax3}
\eea
where $H_n(x)$ is the Hermite polynomial. Notice that $Z_n(x)$
forms a complete set of eigenfunctions of the harmonic oscillator,
\bea
&\Big(\dfrac{d^2}{dx^2} -x^2 \Big) Z_n(x)
=- (2n+1) Z_n(x).
\eea
Substituting (\ref{eqnax3}) into (\ref{1aximon})
we obtain
\bea
&&\sum_{n=0}^{\infty} \Big\{\Big(\dfrac{d^2 f_n}{d\rho^2}
+ \big(\dfrac{1}{\rho}+\dfrac{z^2}{\rho^2} \big)\dfrac{d f_n}{d\rho}
\Big) H_n - \dfrac{z}{\rho \sqrt \rho} \dfrac{df_n}{d\rho}\dfrac{dH_n}{dx}  \nn\\
&&+f_n \Big(\dfrac{z^2}{4 \rho^3} \dfrac{d^2 H_n}{dx^2}
-\dfrac{z}{4\rho^2 \sqrt \rho} \big(1-\dfrac{z^2}{\rho^2}\big)
\dfrac{d H_n}{dx} \Big)  \nn\\
&&+ \Big(\dfrac{z^4}{4 \rho^4}- \dfrac{z^3}{4\rho^3 \sqrt \rho}
+ \dfrac{z^2}{\rho^3} -\dfrac{2n+1\pm2}{\rho} + E \Big) f_n H_n \Big\}\nn\\
&&~~~~~~~~~~~~~~~~~~~~~~~~~~~~~~~~~=0. \label{eqnlong} \eea Using
the recurrence relations and orthogonality properties of Hermite
polynomials one can derive the equations for $f_n(\rho)$ \bea
&&\Big(\dfrac{d^2}{d\rho^2} + \dfrac{2n+3}{2\rho} \dfrac{d}{d\rho}
+\dfrac{4n^2-2n-1}{16 \rho^2} \nn \\
&&-\dfrac{2n+1 \pm 2}{\rho} + E \Big) f_n \nn\\
&&= - \dfrac{1}{64 \rho^2} f_{n-4}+\dfrac{1}{32 \rho^2} f_{n-3}
-\dfrac{1}{4\rho} \Big(\dfrac{d}{d\rho}
+ \dfrac{n-1}{4 \rho} \Big) f_{n-2} \nn \\
&&+\dfrac{3}{16 \rho^2} f_{n-1}
+\dfrac{3(n+1)^2}{8 \rho^2} f_{n+1}  \nn \\
&&+\dfrac{(n+1)(n+2)}{\rho} \Big(\dfrac{d}{d\rho}
-\dfrac{3n+2}{4\rho} \Big) f_{n+2}  \nn \\
&&+\dfrac{(n+1)(n+2)(n+3)}{4\rho^2} \Big(f_{n+3}- 3(n+4) f_{n+4} \Big),
\eea
where $f_n=0$ for negative integer $n$. So we have infinite
number of equations for infinite number of unknown functions $f_n(\rho)$.

Notice that the left hand side of the last equation
contains a second order differential operator with the
quantum mechanical potential
\bea
U = \dfrac{2n+1 \pm 2}{\rho}.
\eea
The potential becomes attractive only if $n=0$, so that
in a first approximation we expect that the negative energy
eigenvalues will originate mainly from the lowest bound state
with $n=0$ of the harmonic oscillator
part. So, by neglecting all $f_n$ with $n\neq 0$
we can still get an approximate qualitative solution
for $f_0$. In such an approximation the equation reduces to
a simple one
\bea
&\Big(\dfrac{d^2}{d\rho^2} + \dfrac{3}{2\rho} \dfrac{d}{d\rho}
+\dfrac{1}{\rho} -\dfrac{1}{16 \rho^2} + E \Big) f_0 =0.
\eea
The solution to this equation has a new integer
quantum number $k$,
\bea
 f_{0,k}(\rho) &=&  \rho^s \exp \big(-\sqrt {|E_k|} \rho \big)
\dfrac{}{} \sum_{l=0}^{l=k} a_l \rho^l, \nn \\
s&=& \dfrac{\sqrt 2 -1}{4}, \nn \\
 E_k &=& -\dfrac{1}{(2k + 2s + 3/2)^2}, \nn \\
 a_{l+1} &=& \dfrac{\sqrt {|E_k|} (2l + 2s + 3/2)-1}{(l+1)
(l+2s + 3/2)} a_l .
\label{eqng0}
\eea
With this we may express the corresponding eigenfunction
$\Psi_k$ as
\bea
& \Psi_k (\rho,\phi,z)= N_k \exp \big(-\dfrac{z^2}{2\rho} \big)
f_{0,k}(\rho),
\label{solution}
\eea
where $N_k$ is a normalization constant. One can find the
lowest energy eigenvalues
\bea
&& E_0 = -0.343..., \nn \\
&& E_1 = -0.073...,\nn \\
&& E_2 = -0.031...,\nn \\
&& E_3 = -0.017..., \nn \\
&& E_4 = -0.011....
\label{spectr}
\eea
This confirms that the axially symmetric monopole string background
is indeed unstable.

Surprisingly, we find that the approximate solution (\ref{solution})
can also be obtained as an exact solution of variational method
with the trial function $\tilde F$ of the form
\bea
\tilde F(\rho,z) = N \rho^s \exp \big(-\beta_k \rho-\gamma
\dfrac{z^2}{2\rho} \big)  \dfrac{}{}\sum_{l=0}^{l=k} a_l \rho^l,
\eea
where $s,\beta_k,\gamma,a_l$ are
treated as variational parameters.
In other words, the variational minimum of the energy functional
with the above trial function is provided exactly
by the solution (\ref{solution}).

To quantify the accuracy of our approximate
analytic solution
we solve numerically the starting equation (\ref{1aximon}) (with the lower
negative sign and $m=0$).
The obtained numerical solution for $\Psi$ have
the same essential singularity structure as in (\ref{solution}).
The corresponding lowest energy eigenvalues
\bea
&& E_0 = -0.545..., \nn \\
&& E_1 = -0.093...,\nn \\
&& E_2 = -0.036...,\nn \\
&& E_3 = -0.019..., \nn \\
&& E_4 = -0.011....
\eea
confirm that the solution (\ref{solution})
provides a good qualitative
estimation of the energy spectrum to analyse the
vacuum stability of the axially symmetric monopole string.
What is more important is that we can apply the structure
of that solution to the analysis of the stability problem
for the monopole-antimonopole string configuration in the subsequent
section.

{\bf 4. A stable magnetic background}

The main idea how to construct a stable magnetic background
is quite clear. Consider a pair of axially symmetric monopole
and anti-monopole strings which are orthogonal
to the $xy$-plane and separated by a distance $a$.
Due to the opposite directions of the magnetic fields
of the monopole and anti-monopole strings
the attractive part of the quantum mechanical potential $U(\rho)$ in the
eigenvalue equation falls down
as $U(\rho) \rightarrow \,\, O(-a/\rho^2)$
when $\rho \rightarrow \infty$. This allows the centrifugal
potential to be competitive with the attractive part
and prevail for small enough values of $a$.
By decreasing the distance $a$ we can decrease
the effective size of the quantum mechanical
potential well, so that the bound state energy levels
will be pushed out from the well.
This, with the positive asymptotics of the potential at infinity,
implies that the bound states will
have disappeared completely at some
finite critical value of $a$.

To show this, consider a pair of axially symmetric monopole and
anti-monopole strings located at $(\rho=a/2,\phi=0)$
and $(\rho=a/2,\phi=\pi)$ in cylindrical coordinates.
The magnetic field strengths  ${\vec H}_{\pm}$ for
the monopole and anti-monopole strings are defined as follows
\bea
 \vec H_{\pm} &=& \pm \dfrac{\alpha}{\rho_{\pm}} \hat \rho_{\pm}, \nn \\
 \vec \rho_{\pm}&=& \vec \rho \pm \dfrac{\vec a}{2}, \nn \\
 \rho_{\pm}^2 &=&\rho^2 \pm a \rho \cos \phi +\dfrac{a^2}{4},
\eea
where $\vec a$ is the two-dimensional vector starting from
the anti-monopole string to the monopole string in $xy$-plane.
From now on we will assume $\alpha=1$ without loss of generality.
The total magnetic field is given by
\bea
\vec H &=& \vec H_+ + \vec H_-, \nn \\
H_{\rho} &=& \dfrac{\rho-\dfrac{a }{2} \cos \phi}{\rho_+^2}
-\dfrac{\rho+ \dfrac{a }{2}\cos \phi}{\rho_-^2}, \nn \\
H_{\phi} &=& \dfrac{a \rho(\rho^2 + \dfrac{a^2}{4}) \sin \phi}
{\rho_+^2 \rho_-^2},
~~~~~H_{z} =0 , \nn \\
H &=& \sqrt {H_\rho^2 + \dfrac{1}{\rho^2}H_\phi^2}
= \dfrac{a}{\rho_+ \rho_-}.
\eea
One can express the corresponding vector potential
in terms of $H_{\rho,\phi}$ components
\bea
&A_\mu = z H_{\phi} \pro_\mu \rho
-\rho z H_{\rho} \pro_\mu \phi.
\eea
The eigenvalue equation for the operator $K$ takes the form
\bea
\Big\{&-&\dfrac{1}{\rho} \pro_\rho (\rho \pro_\rho )
- \dfrac{1}{\rho^2} \pro_\phi^2 - \pro_z^2
- 2i\dfrac{z}{\rho} \Big(H_{\phi} \pro_\rho
-H_{\rho} \pro_\phi \Big) \nn\\
&+& z^2 H^2 \pm 2H \Big\} \Psi(\rho,\phi,z) = E \Psi(\rho,\phi,z).
\label{orig} \eea The equation can be interpreted as a
Schroedinger equation for a massless gluon in the magnetic field
of monopole and anti-monopole string pair.

Let us analyse the equation qualitatively first. We will
concentrate on the potentially dangerous term $-2H$ in
(\ref{orig}). The singularities of the term $z^2 H^2$ determine
the essential singularities of the differential equation. One can
extract the leading factor of the solution and look for a finite
solution for the ground state in the form similar to
(\ref{solution}) \bea \Psi(\rho,\phi,z) = (\pi \rho_+
\rho_-)^{-\frac{1}{4}} \exp\big(-\dfrac{z^2}{2 \rho_+ \rho_-}\big)
F(\rho,\phi), \label{solform} \eea where $F(\rho,\phi)$ is
normalized by \bea \int |F(\rho,\phi)|^2 \rho d\rho   d\phi = 1.
\eea The solution describes a wave function localized mainly near
the string pair. The wave function vanishes exactly on the axes of
the strings. This implies that the ground state has a non-zero
orbital angular momentum which provides a centrifugal potential as
we will see later.

The lowest negative eigenvalue of this equation can be
obtained by variational method by minimizing the corresponding
energy functional
\bea
&E=\dfrac{}{} \int \rho \Psi^* \Big [-\dfrac{1}{\rho} \pro_\rho (\rho \pro_\rho )
-\dfrac{1}{\rho^2} \pro_\phi^2 - \pro_z^2
- 2i\dfrac{z}{\rho} (H_\phi \pro_\rho \nn\\
&- H_\rho \pro_\phi)
+ z^2 H^2 \pm 2H \Big ] \Psi d\rho dz d\phi.
\eea
Now, with
\bea
F(\rho,\phi)= \dfrac{}{} \sum_{-\infty}^{+\infty}
f_m (\rho) \Phi_m (\phi),
\eea
one may suppose that the main contribution
to the ground state energy comes from the first term of Fourier
expansion with $m=0$. With this we can perform the integration
over $z$-coordinate and simplify the above expression to
\bea
&&E=\int \rho f(\rho) \Big(-\pro_{\rho \rho}-\dfrac{1}{\rho} \pro_\rho
+U(\rho,\phi)\Big) f(\rho) d\rho d\phi, \nn\\
&& U(\rho,\phi) = \dfrac{\rho^2 - 2 a \rho_+ \rho_-}{2 \rho_+^2\rho_-^2},
\eea
where $f(\rho)=f_0(\rho)$ and $U(\rho,\phi)$ is an effective
potential. Since the energy
eigenvalues decrease with decreasing the parameter $a$, 
to study the features of the potential
at small $a$ we make the following rescaling
\bea
&&\rho \rightarrow a \rho,
~~~~~~~~~~f \rightarrow f/a,  \nn\\
&&E \rightarrow E/a^2 .
\eea
Under this rescaling the potential near the origin
can be approximated to
\bea
U(\rho,\phi) \rightarrow -4 a
+ (8 -16a \cos 2\phi) \rho^2.
\label{near0}
\eea
So that the potential reduces to a two-dimensional
harmonic oscillator potential whose depth
decreases as $a$ goes to zero.
This implies that the negative energy eigenvalues
can disappear for some small values of $a$
if the asymptotics of the potential at infinity
becomes positive.
To see this we perform
the integration over the angle variable $\phi$
in the energy functional, and with the change of variable
\bea
f(\rho) = \chi (\rho) /\sqrt{\rho},
\eea
we obtain the following equation
which minimizes the energy,
\bea
\Big[-\dfrac{d^2}{d\rho^2}&+&V(\rho) \Big] \chi(\rho) = E \chi(\rho), \nn\\
 V(\rho) &=& -\dfrac{1}{4\rho^2}+ \dfrac{8\rho^2}
{\sqrt{(a^4-16 \rho^4)^2}} \nn \\
&-&\dfrac{8a}{\pi \sqrt {(a^2-4\rho^2)^2 }}
K\big(-\frac{16 a^2 \rho^2}{(a^2 - 4 \rho^2)^2}\big),
\label{eq10}
\eea
where $K(x)$ is the complete elliptic integral of the first kind.
Notice that the first term in the potential does not
produce bound states because the potential $-\dfrac{\kappa}{\rho^2}$
leads to negative energy eigenvalues only for
$\kappa >1/4$.
From the last equation we obtain the asymptotic behavior of
the potential $V(\rho)$ near space infinity
\bea
V(\rho) \simeq (\dfrac{1}{4}-a) \dfrac{1}{\rho^2}.
\eea
This tells that the potential becomes positive when the distance
$a$ becomes less than the critical value $a_{cr}$ (in the unit $1/\alpha$)
\bea
a<a_{cr}=\dfrac{1}{4}.
\eea

To check the analytic estimate of the critical value $a_{cr}$
we solve numerically the original equation (\ref{orig}).
Since the wave function $\Psi(\rho,\phi,z)$ becomes spread
to long distance for small energy eigenvalues
we extend sufficiently the upper limit of $\rho$
in the domain $(0<\rho<\rho_{upper}, 0<\phi<2 \pi)$
and increase the mesh near the location of the
monopole and antimonopole strings.
The extrapolation of the results from finite
small energy eigenvalues to zero
gives the following critical value within 6\% of accuracy
\bea
a_{cr}\simeq 0.246 \, .
\eea
The numerical result confirms that qualitatively the
approximate solution (\ref{solform})
describes the correct physical picture. In particular,
this tells that a pair of monopole and antimonopole strings
becomes a stable magnetic background if the distance between
two strings is small enough.

{\bf 5. On stability of vortex-antivortex pair}

Recently an alternative mechanism of confinement has been proposed
which advocates the condensation of magnetic vortices \cite{diak}.
However, it has been known
that the magnetic vortex configuration
is unstable \cite{bordag}. So it would be interesting to
study the stability of the vortex-antivortex pair.
In this section we study the stability of
a special vortex-antivortex configuration.

Let us start with a single magnetic vortex given by
\bea
&\vec H= \dfrac{1}{\rho} \hat z,
~~~~~A_\mu =\rho \pro_\mu \phi.
\eea
Notice that, unlike the monopole string (\ref{mstring}),
the vortex configuration is not a classical solution of the system.
But since such a type of configuration multiplied by
an appropriate profile function has been studied
in the framework of various approaches, we will
consider that special vortex configuration in the following.

One can write down the corresponding
eigenvalue equation of the operator $K$
\bea
&\Big[-\pro_\rho^2-\dfrac{1}{\rho} \pro_\rho - \dfrac{1}{\rho^2} \pro_\phi^2
-\pro_z^2 -\dfrac{2i}{\rho} \pro_\phi
\pm 2 H \Big] \Psi(\rho,\phi,z) \nn\\
&= E \Psi(\rho,\phi,z).
\label{schr3}
\eea
The equation becomes separable in all three variables.
With factorization
\bea
&\Psi= \dfrac{}{} \sum_{-\infty}^{+\infty}
f_m(\rho) g(z) \Phi_m(\phi),~~~~~g(z)=1,
\eea
one obtains the following ordinary differential
equation for $f(\rho)$ from (\ref{schr3}),
\bea
\Big(-\pro_\rho^2-\dfrac{1}{\rho} \pro_\rho
+ (1+\dfrac{m}{\rho})^2 \pm \dfrac{2}{\rho} - E\Big) f(\rho) =0.
\eea
The bound states are possible for the potential $-2/\rho$
with non-positive integer $m$,
in which case the corresponding solution
can be obtained
\bea
&& f_{n,m}(\rho) = \rho^{|m|} e^{-\sqrt{1-E_{n,m}}} u_{n,m}(\rho), \nn \\
&& u_{n,m} (\rho) = \dfrac{}{}\sum_{k=0}^n a^{n,m}_k \rho^k, \nn \\
&& a^{n,m}_{k+1} =
\dfrac{\sqrt{1-E_{n,m}} (2 k +2 |m|+1) -2+2m}
{(k+1)(k+2 |m| +1)} a^{n,m}_k, \nn \\
&& E_{n,m} = 1-\dfrac{4 (1-m)^2}{(2n+2 |m| +1)^2}, \nn \\
&& n=0,1,2,...; \,\,\, m=0,-1,-2,...
\eea
Clearly, the ground state has a negative energy $E_{0,0}$,
which tells that the vortex configuration is unstable.

There is a principal difference between the axially symmetric 
monopole string and the vortex configuration. 
The ground state of the monopole string
has a non-trivial centrifugal potential. 
But the ground state eigenfunction $f_{0,0}(\rho)$ of the 
vortex configuration corresponds to an $S$-state,
which implies the absence of the centrifugal potential.
This plays a crucial role in the existence of the negative 
energy eigenstates in the case of vortex-antivortex pair.

The vortex-antivortex background is described in
a similar manner as the monopole-antimonopole string
background in the previous section. The potential has the form
\bea
A_\mu &=& \dfrac{a}{2} \sin \phi
(\dfrac{1}{\rho_+}+\dfrac{1}{\rho_-}) \pro_\mu \rho \nn \\
&&+ \Big[\dfrac{\rho}{\rho_+} (\rho+\dfrac{a}{2} \cos \phi)
-\dfrac{\rho}{\rho_-}(\rho-\dfrac{a}{2} \cos \phi) \Big] \pro_\mu \phi, \nn\\
 \vec H &=& \Big(\dfrac{1}{\rho_+} - \dfrac{1}{\rho_-} \Big) \hat z,
\eea
where $a$ is the distance between the axes of the vortex and anti-vortex.
The eigenvalue equation corresponding to the
operator $K$ is partially factorizable in $z$-coordinate
\bea
&\Big[-\pro_\rho^2-\dfrac{1}{\rho} \pro_\rho
-\dfrac{1}{\rho^2} \pro^2_\phi-\pro_z^2
- 2i (A_\rho \pro_\rho+\dfrac{1}{\rho^2} A_\phi \pro_\phi)\nn \\
&+A_\mu^2 \pm 2H \Big] F(\rho,\phi,z)=E F(\rho,\phi,z) .
\eea

The numerical analysis of the equation shows that there is no
critical value for the parameter $a$, so that the negative energy
eigenvalues exist for any small $a$. Qualitatively one can see
this from the effective potential $V = A_\mu^2-2H$. After
averaging over the angle variable one can find the asymptotic
expansion of the potential near the origin and infinity 
\bea &
V(\rho)\simeq \left\{{\dfrac{}{}8\pi-\dfrac{64}{a^2} \rho -
\dfrac{16 \pi}{a^2} \rho^2~~~(\rho \simeq 0), \atop
~-\big(\dfrac{}{}4-\dfrac{\pi}{2}a \big) \dfrac{2a}{\rho^2}
~~~~~~~(\rho \simeq \infty).}\right. 
\eea 
This shows that there is no centrifugal term which could
prevent the appearance of bound states for small $a$. Whether the
instability problem can be overcome with a more complicate
configuration of the vortex-antivortex remains an open question.

In conclusion, we have shown that the axially symmetric
monopole-antimonopole string background is stable under the small
field fluctuation if the distance between two strings becomes less
than the critical value $a_{cr} \simeq 1/4$. The existence of the
stable classical magnetic background implies that ``a spaghetti of
monopole-antimonopole string pairs" could generate a stable vacuum
condensation. This would allow a magnetic confinement of
color in QCD. 

{\bf Acknowledgements}

One of the authors (YMC) thanks G. Sterman for the kind
hospitality during his visit to YITP. Another author (DGP) thanks
K.-I. Kondo and N. I. Kochelev for useful discussions. The work is
supported in part by the ABRL Program of KOSEF
(R14-2003-012-01002-0) and by the BK21 Project of the Ministry of
Education.


\begin{thebibliography}{99}
\bibitem{nambu} Y. Nambu, Phys. Rev. {\bf D10} (1974) 4262;
S. Mandelstam, Phys. Rep. {\bf 23C} (1976) 245;
A. Polyakov, Nucl. Phys. {\bf B120}  (1977) 429;
G. 't Hooft, Nucl. Phys. {\bf B190} (1981) 455.
\bibitem{cho1} Y. M. Cho, Phys. Rev. {\bf D21}  (1980) 1080;
Phys. Rev. Lett. {\bf 46} (1981) 302;
Phys. Rev. {\bf D23} (1981) 2415; Z. Ezawa and A. Iwazaki,
Phys. Rev. {\bf D25} (1982) 2681.
\bibitem{savv}  G. K. Savvidy, Phys. Lett. {\bf B71} (1977) 133;
N. K. Nielsen and P. Olesen, Nucl. Phys. {\bf B144} (1978) 376;
\bibitem{cox}  A. Yildiz and P. Cox, Phys. Rev. {\bf D21} (1980) 1095;
M. Claudson, A. Yilditz, and P. Cox, Phys. Rev. {\bf D22} (1980) 2022;
W. Dittrich and M. Reuter, Phys. Lett. {\bf B128} (1983) 321;
M. Reuter, M. G. Schmidt, and C. Schubert,
Ann. Phys. {\bf 259} (1997) 313; C. Flory, Phys. Rev. {\bf D28} (1983) 1425;
S. K. Blau, M. Visser, and A. Wipf, Int. J. Mod. Phys.
{\bf A6} (1991) 5409; V. Schanbacher, Phys. Rev. {\bf D26} (1982) 489.
\bibitem{niel2} H. B. Nielsen and M. Ninomiya,
Nucl. Phys. {\bf B156} (1979) 1;
N. K. Nielsen and P. Olesen, Nucl. Phys. {\bf B160} (1979) 380.
\bibitem{rajiad} H. Leutwyler, Nucl.Phys. {\bf B179} (1981) 129; Phys. Lett.
{\bf B96} (1980) 154; C. Rajiadakos, Phys. Lett. {\bf B100} (1981) 471.
\bibitem{cho3}  Y. M. Cho, H. W. Lee, and D. G. Pak,
Phys. Lett. {\bf B 525} (2002) 347; Y. M. Cho and D. G. Pak,
Phys. Rev. {\bf D65} (2002) 074027; Y. M. Cho and M. L. Walker,
Mod. Phys. Lett. {\bf A19}, 2707 (2004);
K.-I. Kondo, Phys.Lett. {\bf B600} (2004) 287.
\bibitem{kondo1}  S. Kato, K.-I. Kondo, T. Murakami,
A. Shibata, T. Shinohara, hep-ph/0504054.
\bibitem{brandt} R. A. Brandt and F. Neri, Nucl. Phys. {\bf B 161}
(1979) 253.
\bibitem{wu1} T. T. Wu and C. N. Yang, Phys. Rev. {\bf D12} (1975) 3845.
\bibitem{cho80} Y. M. Cho, Phys. Rev. Lett. {\bf 44} (1980) 1115;
Phys. Lett. {\bf B115} (1982) 125.
\bibitem{wu2} T. T. Wu and C. N. Yang, Nucl. Phys. {\bf B 107} (1976) 365;
Phys. Rev. {\bf D 16} (1977) 1018.
\bibitem{landau}  L. D. Landau and E. M. Lifshitz,
{\it Course of Theoretical Physics: Vol. 3, Quantum Mechanics},
(Pergamon Press, 1977).
\bibitem{diak}  D. Diakonov and M. Maul, Phys. Rev. {\bf D66} (2002) 096004;
J. D. Lange, M. Engelhardt and H. Reinhardt, Phys. Rev. {\bf D68}
(2003) 025001.
\bibitem{bordag} M. Bordag, Phys. Rev. {\bf D67} (2003) 065001.
\end{thebibliography}
\end{document}